\documentclass{article}
\usepackage[utf8]{inputenc}
\usepackage{glossaries}
\usepackage{booktabs}
\usepackage{array}
\usepackage{amsmath}
\usepackage{graphicx}
\usepackage{authblk}
\usepackage{tikz}
\def\checkmark{\tikz\fill[scale=0.4](0,.35) -- (.25,0) -- (1,.7) -- (.25,.15) -- cycle;} 

\newacronym{wds}{WDS}{water distribution systems}
\newacronym{spot}{TEVA-SPOT}{threat ensembre vulnerability assesment and sensor placement optimization tool}

\title{Sensor Placement for Contamination Detection in Water Distribution Systems}
\author{Margarita Rebolledo}
\author{Sowmya Chandrasekaran}
\author{Thomas Bartz-Beielstein}
\affil{Institute for Data Science, Engineering, and Analytics, TH K\"oln. Germany}

\begin{document}

\maketitle

\abstract{Sensor placement for contaminant detection in water distribution systems (WDS) has become a topic of great interest aiming to secure a population's water supply. Several approaches can be found in the literature with differences ranging from the objective selected to optimize to the methods implemented to solve the optimization problem. In this work we aim to give an overview of the current work in sensor placement with focus on contaminant detection for WDS.  We present some of the objectives for which the sensor placement problem is defined along with common optimization algorithms and Toolkits available to help with algorithm testing and comparison. }

\section{Introduction}
Sensor placement is a topical issue in industry with important applications in \gls{wds}. The terrorist attacks from 2001 have been a constant force driving the efforts to minimize the vulnerability of \gls{wds} to natural and terrorist threats \cite{EPA11a}. A typical concern is the insertion of biological or chemical agents into the water supply. For this an early warning detection system is needed to quickly identify such an attack.\\

To reduce the vulnerability of \gls{wds} and control any degradation in water quality a reliable sensor network is required. However, given the extensiveness of the water networks the costs associated with the installation, material and data managing of the monitoring system are not negligible. 

Recent literature has focused on finding the optimal layout of sensors for early warning systems. This is not a straightforward task given the different and opposing goals, such as minimizing costs while maximizing the coverage of the sensor network. Another factor comes form the different definitions of the intended final goal, minimizing the affected population or the total volume of contaminated water.\\

The main aim of this work is to give an overview of  sensor placement focusing on contaminant detection for \gls{wds}. We highlight some of the used approaches and tools available.

The paper is structured as follows, first in Section \ref{sec:objectives} different objectives found in the bibliography to define an early warning system are presented. In Section \ref{sec:optimization} a short overview of the optimization algorithms and methods used to solve the different design problems are given. Section \ref{sec:toolkits} list the toolkits available to help in the definition, solution, and comparison of sensor placement problem. Lastly, Section \ref{sec:diss} gives final remarks.

\section{Sensor Placement Objectives} \label{sec:objectives}
There are several competing objectives when defining the sensor placement problem. Some examples include the minimization of the time of detection for a certain contaminant, maximizing the network coverage or the probability of detection. 
In the case of contaminant detection it is generally accepted that the impact on the public should remain as low as possible.
This involves above other metrics to keep the number of exposed people to a minimum, maintaining the exposition to the contaminant under a certain threshold or minimizing the volume of contaminated water.\\

The objective or objectives selection will influence the final sensor placement. Results obtained after the Battle of the Water Sensor Networks \cite{Ostf08a}, hosted on the 8th Annual Water Distribution Systems Analysis Symposium (Cincinnati, Ohio, August 27–29, 2006), reached the conclusion that no single sensor placement formulation or optimization solving method is superior. 
In the following some of the implemented sensor placement objectives found in the bibliography are presented: \\

Sensor connectivity problems can occur on \gls{wds} with difficult topography or difficult sensor locations, as is the case with underground sensors.
In \cite{Bert14a} a sensor placement strategy is studied in which the objective is to maximize the sensors' probability of detection while simultaneously ensuring the communication between sensors is maintained. 
The proposed optimization problem is defined as a integer programming problem constrained by the sensors ability to communicate among themselves. The method is tested on three different \gls{wds} simulations and found to increase the probability of detection in the sensor network.

The work of \cite{Aral10a} proposed a formulation to include four different competing criteria into a single objective function. The formulations goal is to find and generate a balance between all the objectives without external pressure, ensuring that the problem formulation is kept as single objective. The formulation combined: (1) the expected time of detection, (2) the expected population affected prior the time of detection, (3) the volume of consumed contaminated water prior to detection, and (4) the reliability of the sensors. The first three of the objectives are to be minimized and the last to be maximized respectively. 
The approach was tested on two differently sized simulated water networks and solutions proved that no objective was emphasized over the other. Additionally, when compared to multi objective formulations the proposed methods required less computational effort.

In \cite{Pill15a} the authors propose an early warning detection system to optimize the average time to detection, the fraction of population exposed, the likelihood of detection, the average fraction of population exposed at risk, and the installation cost. The sensor design is defined as a Nonlinear Integer Programming problem that is multi-objective. This problem is solved using a customized greedy algorithm. The approach was tested on the Communaut\'{e} urbaine de Strasbourg (Strasbourg Eurom\'{e}tropole) water network. Results showed that no more than 94 out of 200 additional sensor were needed for a 95\% detection of contamination events in less than 5 hours.

As a final example, authors in  \cite{Ung17a} focus on solving the sensor placement problem that optimizes the results of source identification methods. It defines one criterion which it names \textit{Contribution} as the optimization objective. The \textit{Contribution} takes into consideration the detection likelihood, accuracy and specificity of the source identification method on all the simulated contamination scenarios. The authors proposed source identification method uses a backtracking algorithm to identify the node that most probably activated a sensor reading. After the source identification is executed its \textit{Contribution} can be computed. The proposed methodology is tested on a real french water network. The results showed that solving the optimization using a greedy algorithm produced the best performance overall but at a high time and computational cost. On the other hand, using a local search algorithm lowered the performance but increased the time efficiency.\\

More implementations of sensor placement with single and multi objective formulations can be found in \cite{Rathi14a, Aded19a}

\section{Optimization Methods} \label{sec:optimization}
The sensor placement problem can be formulated as an optimization problem. According to the implemented formulation several optimizers
are available to solve the defined problem. As already presented in some of the objective formulations described in Section \ref{sec:objectives}, integer programming solvers, heuristics or local search algorithms can be used to find an optimal solution to the sensor placement problem.
 
Run-time, performance or computational efficiency are some of the most common factors differentiating them and form the base for selecting the appropriate one.

Greedy algorithms were used on \cite{Pill15a} \cite{Ung17a} and \cite{Berr06a} to solve the single objective optimization problem.

In the method described by \cite{Aral10a}, the optimization problem is described as an \{0,1\} integer programming, where the \gls{wds} nodes are represented as 0 if no sensor is placed and 1 otherwise. To reduce computational costs and allow for large network inclusion, a progressive genetic algorithm was used to solve the optimization problem.

\cite{Hu15a} Proposed a co-evolution approach to tackle the sensor placement problem. In this framework, multiple population subgroups cooperate to find an optimal solution. Each population is evolved independently with a step to exchange information after every iteration. For the studied case, several particle swarm populations were tested against a genetic algorithm and a particle swarm algorithm. The use of the proposed algorithm was able to reduce the time of detection of a contaminant in the \gls{wds}.

An algorithm derived from the cross entropy method, a Monte Carlo method, was implemented by \cite{Dori06a} to solve a multi-objective sensor placement problem. 
In their approach the authors modified the algorithm by adding a two stage process and an additive noise disturbance to the updating mechanism. The derived algorithm is referred as Noisy Cross Entropy Sensor Locator (nCESL) algorithm. The algorithm was applied to solve four conflicting objectives:  (1) minimisation of the expected time of detection, (2) minimisation of the expected population affected prior to detection, (3) minimisation of the expected demand of contaminated water prior to detection and (4) maximisation of the detection likelihood.

\section{Toolkits} \label{sec:toolkits}
Sensor placement software tools are important to compare sensor placement methods and allow reproducible research. An overview of some of the software tools available to test contaminant warning systems in \gls{wds} is discussed in this section.\\

EPANET \cite{Ross00a} is a public domain software used to model water distribution systems. It contains a state-of-the-art hydraulic analysis engine that allows to perform hydraulic and water quality behaviour simulations within pressurized pipe networks. Among other functionalities, EPANET is able to track the flow of water at each pipe, the chemical concentrations, the pressure at nodes and the source of an event throughout a simulation period. 
EPANET has been used to conduct consumer exposure assessments, vulnerability studies and to evaluate strategies for improving water quality. \cite{Aral10a,Bert14a,Dori06a} Used the EPANET software to test their sensor placement approaches.\\

The \gls{spot} \cite{Hart08a} allows to develop sensor network designs for large water utilities. It allows to (1) simulate contamination incidents, (2) compute contamination impacts, (3) perform sensor placement, and (4) evaluate a sensor placement. In the first step of the \gls{spot} process EPANET is employed to simulate a set of contamination incidents. Subsequently the impact of the incidents is calculated with respect to specified objectives e.g. extent of contamination or time until detection. The sensor placement is performed on the basis of the generated impact. 
This is done following the standard SPOT formulation which is to minimize the expected impact of a group of contamination incidents given a fixed sensor budget. The expected impact a sensor placement configuration has is formulated as a mixed integer problem taking into account constraints and response times. \gls{spot} offers interfaces to known solvers in order to solve the impact
formulation: Mixed integer programming solvers C-PLEX and PICO, GRASP heuristic and Lagrangian heuristic. 
\gls{spot} is no longer on active development with version 2.5.2 being the final release.\\

Chama \cite{Klis17a} is an open source Python package designed to be a general purpose sensor placement optimization software tool. It includes some of the methods previously developed for \gls{spot}. As in \gls{spot}, Chama defines the sensor placement problem as a mixed integer programming problem. This problem formulation can be used to minimize impact or maximize coverage.
Chama methods are currently used in contamination warning systems, but are general and can be applied to a wide range of applications. Its usage includes five steps that are (1) simulation, (2) sensor technology, (3) impact assessment, (4) optimization and (5) graphics. The workflow is modular and the user can enter it at any given time.
The simulation data can be directly given into Chama or read from water network simulators as EPANET or WNTR \cite{Klis17b}.  
An improvement from \gls{spot} is the inclusion of a sensor module which allows to define sensor properties such as position -mobile or stationary-, detector type -point or camera-, sampling time or detection threshold.
The impact is calculated with respect to specified objectives, such as time to detection or population impacted. The formulation is equivalent to \gls{spot} standard formulation. Additionally, the impact metrics can be used for coverage-based optimization formulations.
The optimization formulation for either the impact or the coverage formulation is defined and solved through Pyomo \cite{Hart12a}. The default solver is GLPK but other MIP solvers can also be used. \\

The sensor placement toolkit (S-PLACE) \cite{Elia14a} is a software for the development and comparison of different sensor placement methods and algorithms implemented in Matlab's programming language. It is based on the EPANET-Matlab toolbox which wraps all the EPANET functionalities. 
S-PLACE has a modular design that allows an easy customization. In the 'Data Module' the EPANET model is constructed. The 'Scenarios Contruction Module' allows to specify the contamination scenarios and simulate them in EPANET. The simulated damage generated by each scenario is calculated in the 'Impact matrix calculation module'. Finally, the 'Sensor Placement Module' computes the final solutions using the solver specified by the user. 
The optimization problem is formulated as a multi-objective risk-minimization problem with four optimization functions: (1) the number of sensors, (2) the average impact risk, and (3) the estimated worst case impact.\\

The Online Security Management and Reliability Toolkit for Water Distribution Networks project \cite{smartwdn} (SMaRT-Online$^\mathit{WDN}$) is the result of a French-German cooperative research project funded by the ANR (reference project: ANR-11-SECU-006) and the BMBF (reference project: 13N12180). 
The main objective of SMaRT-Online$^{WDN}$ is the development of a security management toolkit for water distribution networks based on sensor measures of water quality and quantity. 
Among the research objectives of this project is the optimal location of sensors. The main difference to other existing sensor placement tools is the reformulation of the problem as an online hydraulic and water quality monitoring system. This reformulation can overcome the data uncertainty and random variations on the network product of unknown flow conditions at the moment a contaminant is detected in the network. \\

\begin{table}
\caption{Overview on the characteristics of the described Toolkits. A checkmark indicates the feature is available. An asterisk (*) indicates there is no information on this feature.\\}
\label{tab:sum}
	\centering
	\renewcommand{\arraystretch}{1.5}
	\resizebox{\textwidth}{!}{
	\begin{tabular} {|l|c|c|c|c|c|} 
		\hline
		Feature & EPANET  & \gls{spot} & Chama & S-PLACE & SMaRT-Online$^\mathit{WDN}$\\
		\hline
		Hydraulic modeling & \checkmark & & & & \checkmark\\
		Online & & & & & \checkmark\\
		Open Source &\checkmark &\checkmark &\checkmark & & \\
		Modular & & & \checkmark&\checkmark &\checkmark \\
		Sensor specification & & \checkmark&\checkmark &\checkmark & \\
		Environment & exec & exec & Phyton & Matlab & *\\
		Flexibility & & &\checkmark &\checkmark &\\
		Generalizability &\checkmark & &\checkmark &\checkmark & *\\
		\hline 
	\end{tabular}}
\end{table}

A summary of the characteristics of the presented toolkits is given in Table \ref{tab:sum}. The flexibility is defined as the ability to enable adaptations or extension to the normal system process. This allows quick changes and variations to fit several user needs. Toolkits implemented in common modular programming languages, like Python and Matlab, are easy to modify given its programming language extensive use and material availability.
Generalizability is taken as the ability to support many use cases without extensive changes or large susceptibility to error. One simple example of generalizability is the use of different metric systems which can lead to failures and errors in the calculations. EPANET allows both US and international metrics systems according to desired output measures. Here care should be taken when defining the network to not confuse both metrics. Software like Chama and S-PLACE avoid this source of confusion by avoiding absolute measures and only using unitless values relative to the desired users metrics.

\section{Discussion and Closing Remarks} \label{sec:diss}
Sensor placement for \gls{wds} is not an unified problem design. It depends on the definition of several and different optimization goals. No single definition or solving methods has been proved to be the best option among all.
The final objective, the size and topography of the water network, time and sensor costs determine the problem definition and posterior solving strategies.

Several open source toolkits have been designed in order to allow an easy problem definition and allow for reproducible research. These toolkits, however, are still constrained to fixed objectives definitions and solving strategies leaving room open for improvements.

\section*{Acknowledgements}
This research work is funded by project \textit{Open Water Open Source(OWOS)} (reference number: 005-1703-0011) and is kindly supported by the FH Zeit f{\"u}r  Forschung, Ministerium fur Innovation, Wissenschaft und Forschung des Landes NRW.
 \\

\bibliographystyle{plain}
\bibliography{owos_sensor}
\end{document}